\documentclass[amsmath,amssymb,nofootinbib,showpacs,twocolumn,nofootinbib]{revtex4-1} 
\usepackage{dcolumn}
\usepackage{bm}
\usepackage{graphicx}
\usepackage{epsfig}
\usepackage{color}
\usepackage{wasysym}
\usepackage{natbib}
\usepackage{twoopt}
\usepackage{dashrule}

\definecolor{slateblue}{rgb}{0.2,0.2,0.6}
\usepackage[breaklinks=true,colorlinks=true,linkcolor=slateblue,citecolor=slateblue,urlcolor=slateblue,pdfauthor={Tomassetti},pdftitle={LagVSCycle2022}]{hyperref}

\usepackage{xspace}
\newcommand{\ApJ}{ApJ}

\newcommand{\PRL}{PRL}
\newcommand{\PRD}{PRD}
\newcommand{\ASR}{ASR}
\newcommand{\JGR}{JGR}

\newcommand{\GCR}{CR\xspace}
\newcommand{\GCRs}{CRs\xspace}
\newcommand{\AMS}{\textrm{AMS}\xspace}
\newcommand{\etal}{et al.}
\newcommand{\ie}{\textit{i.e.}} 
 
\def\citep#1{\cite{#1}}

\begin{document}
\title{Temporal evolution and rigidity dependence of the\\ solar modulation lag of Galactic cosmic rays}
\author{Nicola Tomassetti, Bruna Bertucci, Emanuele Fiandrini}
\address{Dipartimento di Fisica e Geologia, Universit{\`a} degli Studi di Perugia, Italy} 

\begin{abstract} 
  When traveling in the heliosphere, Galactic cosmic rays (\GCRs) are subjected to the solar modulation effect,
  a quasiperiodical change of their intensity caused by the 11-year cycle of solar activity.
  Here we investigate the association of solar activity and cosmic radiation over five solar cycles,
  from 1965 to 2020, using a collection of multichannel data from neutron monitors, space missions, and solar observatories. 
  In particular, we focus on the time lag between the monthly sunspot number and the \GCR flux variations.
  We show that the modulation lag is subjected to a 22-year periodical variation, ranging from about 2 to 14 months
  and following the polarity cycle of the Sun's magnetic field.  
  We also show that the lag is remarkably decreasing with increasing energy of the \GCR particles. 
  These results reflect the interplay of basic physics phenomena that cause the \GCR modulation effect:  
  the drift motion of charged particles in the interplanetary magnetic field,
  the  latitudinal dependence of the solar wind, the energy dependence of their residence time in the heliosphere.
  Based on this interpretation, we end up with a global effective formula for the modulation lag
  and testable predictions for the flux evolution of cosmic particles and antiparticles over the solar cycle.
\end{abstract}
\maketitle

\section{Introduction}    
\label{Sec::Introduction} 
%
When traveling inside the heliosphere, Galactic cosmic rays  (\GCRs) interact with
magnetic fields and solar wind disturbances, which cause variations in their intensity and energy spectrum. 
This phenomenon is known as \emph{solar modulation} of \GCRs and it is crucial for the investigation of the
origin of cosmic particles and antiparticles \citep{Potgieter2013}. Understanding \GCR modulation is also
important for assessing the radiation dose and risk in manned space missions \citep{DuranteCucinotta2011}.
In this respect, there are strong efforts aimed at predicting the evolution of the \GCR intensity
near-Earth or in the interplanetary space \citep{Norbury2018}. 
An important feature of solar modulation is its connection with the 11-year cycle of solar activity.
The monthly \emph{sunspot number} (SSN), the main indicator of solar cycle, is known to be anticorrelated with
the long-term variations of the \GCR flux \citep{CletteLefevre2016,VanAllen2000,Usoskin2011}. 
During periods of solar cycle minimum, the flux of Galactic CRs in the inner heliosphere is more intense.
During periods of solar cycle maximum, \GCRs are shielded more effectively by the Sun.
Other proxies for solar activity include the intensity of the solar magnetic field $B_{0}$,
its open magnetic flux, or the tilt angle of the heliospheric current sheet.

Solar activity is constantly monitored by ground based observatories or space probes. The time dependence of the \GCR
flux is measured by several experiments. Direct measurements of particle- and energy-resolved \GCR flux have
been done in space by long-running experiments such as the MED instrument on IMP-8 (since 1972 to 2000 \citep{McDonald2003}),
the HET telescopes on the Voyager probes (1979-present \citep{Cummings2016}), the spectrometers
CRIS on ACE (1997-present \citep{Wiedenbeck2009}), EPHIN on SOHO (1995-present \citep{Kuhl2016}).
Recent measurements include the spectrometers PAMELA on the Resurs-DK1 satellite (2006-2016 \citep{Adriani2013,Martucci2018})
and \AMS on the International Space Station (2011-present \citep{Aguilar2018PHe,Aguilar2021Report}).
Indirect measurements of the \GCR time dependence are performed continuously, since the 1950s,
by neutron monitors (NMs) \citep{Vaisanen2021}. 
NMs show excellent time resolution and exposure, but they have no particle or energy resolution capabilities.

To understand the dynamics of \GCR modulation and its association with solar activity,
it is important to study the \emph{time lag} between the two phenomena.
Several studies reported a lag of few months between the monthly SSN and the corresponding variations in the NM rates
\citep{Nymmik2000,Usoskin2001,SingSingBadruddin2008,Kane2014,AslamBadruddin2015,Chowdhury2016,RossChaplin2019,Iskra2019,Koldobskiy2022TimeLag}. 
Using direct \GCR measurements from space experiments, a lag of 8.1$\pm$0.1 months was reported for Solar Cycle 23 \citep{Tomassetti2017TimeLag}. 
Similar lags were also observed using different proxies such as tilt angle or magnetic field \citep{MishraMishra2018}. 
Such a lag in \GCR modulation is usually interpreted in terms of the plasma dynamics \citep{Nagashima1980,Xanthakis1981,Mavromichalaki1984}.
In fact, the time dependence of \GCRs near-Earth is linked to their transport through the expanding heliosphere.
To first approximation, the lag between \GCRs and solar activity reflects the time spent by the heliospheric
plasma to travel from the Sun to the whole modulation region, which is of the order of one year. 
Other interpretations of the lag include a delayed response of the \GCRs to changes in the background plasma,
or delayed formation of the solar magnetic field with respect to sunspots \citep{DormanDorman1967,VanAllen2000,Wang2022TimeLag}.
On the lag values reported in different analysis there is no clear consensus,
as they range from 0 to 18 months depending on epochs, cycles, NM stations, or indicators \citep{RossChaplin2019,Iskra2019,Nymmik2000}.
Previous studies reported a remarkable odd-even dependence of the lag in terms of cycle number. This effect 
may be ascribed to the role of drift in the heliospheric modulation process \citep{Usoskin2001}.

In this paper, we investigate the delayed relationship between solar activity and \GCR flux over five solar cycles
between 1965 and 2020, \ie, from Cycles 20 to 25. 
In Sec.\,\ref{Sec::Calculations}, we present the theoretical framework to analyze the long-term \GCR modulation using NM and spacecraft data. 
In Sec.\,\ref{Sec::Results}, we present the main results of our correlation analysis. 
In particular, we present a reconstruction of the \GCR modulation lag over the solar cycle.
We show that the lag is subjected to a quasiperiodical evolution, following the 22-year cycle of magnetic polarity,
while its average value is found to decrease with the increasing rigidity (or energy) of the cosmic particles. 
Based on these findings, we also provide an effective formula to describe the
temporal evolution and the rigidity dependence of the \GCR modulation lag. This constitute an essential input
for developing predictive models of \GCR modulation \citep{Slaba2020,Kuznetsov2017,Matthia2013,ONeill2010,Adams2011,ZhaoQin2013}. 
In Sec.\,\ref{Sec::Interpretation}, we discuss the astrophysical interpretation of our findings.
We argue that the two features can be interpreted as signatures of charge-sign dependent drift and energy-dependent
diffusion of \GCRs in the heliosphere, respectively.
This may offer a new way to investigate basic plasma astrophysics processes in the heliosphere. 
Based on this interpretation, we also provide testable predictions using \GCR flux measurements of particles and antiparticles.

\section{Calculations}    
\label{Sec::Calculations} 
%
Our work is based on three main sets of data organized in forms of time series:
direct measurements of \GCR fluxes performed in space, 
counting rates of secondary particles recorded by NMs, and the monthly SSN recorded in solar observatories. 
In this section, we present our calculation framework. First, we briefly outline our calculations for the local interstellar spectra (LIS) of the main \GCR species.
Then, we present the solar modulation calculations used to describe the temporal dependence of the top-of-atmosphere (TOA) fluxes.
Finally, we present the modeling of the counting rate of ground NM detectors.

\subsection{Modeling the LIS fluxes} 
\label{Sec::LISModel}                
%
In this work, LIS fluxes are used as a mere input for the heliospheric modulation calculations.
The investigation of the \GCR propagation in interstellar space is beyond the scope of this work. 
Nonetheless, we opted for fully numerical calculations that incorporates the essential physics of interstellar \GCR transport.
Such an approach allows for a robust estimation of the LIS model uncertainties, that
are considerable in the GeV energy region where no direct data are available \citep{Tomassetti2018PHeVSTime,Tomassetti2019Numerical}.
The relevant LISs for this work are the most abundant species such as \GCR proton ($\sim$\,90\,\%) and helium nuclei
($\sim$\,9\,\%). They are predominantly of primary origin, \ie, accelerated by nonthermal processes in
supernova shockwaves or stellar winds. Other species such as C-N-O nuclei or rarer elements are of minor relevance.
To constrain the Galactic propagation parameters, however, calculations of secondary species
such as lithium or boron are of fundamental importance. 
In our calculations, we implemented a \emph{two-halo model} of \GCR propagation in the Galaxy \citep{Tomassetti2015TwoHalo,Feng2016}.
In this model, the transport of \GCRs in the ISM is described by spatial dependent diffusion of \GCRs in a two-zone magnetic turbulence,
along with their interactions with the gas in the galactic disk.
The acceleration of primary \GCRs is described by rigidity-dependent 
source functions of the type $S_{\rm p}\propto(R/{\rm GV})^{-\nu}$, with index $\nu=$\,2.28$\pm$0.12 for $Z=1$ and index $\nu=$\,2.35$\pm$0.13 for all $Z>1$ nuclei.
The production of secondary particles is calculated using source functions of the type $S_{\rm s}= \sum_{\rm h} \Gamma_{h{\rightarrow}s}^{\rm sp} N_{\rm h}$,
which describe the $h\rightarrow{s}$ fragmentation of $h$-type species of density $N_{\rm{h}}$ into $s$-type nuclei at rate $\Gamma_{h{\rightarrow}s}$.
In this description, the physics processes of acceleration, nuclear fragmentation and ionization losses occur in the Galactic disk,
\ie, where the sources and the matter are placed. The diffusive transport of \GCRs takes place in an extended halo of vertical size $L$.
The \GCR diffusion coefficient in the Galaxy is expressed as $D \equiv \beta D_{0}(R/GV)^{\delta_{i/o}}$,
where we assume two diffusion regimes: a shallow diffusion for the near-disk region with index ${\delta_{i}}$ within distance $z{\leq}\xi{L}$ from the disk,
and a faster diffusion in the extended halo with index ${\delta_{o}}\equiv\delta_{\i}+\Delta$.
Based on our fits, we adopt $\delta_{i}=0.18\pm$0.05 and  $D_{0}/L=0.01\pm$0.002\,kpc/Myr, $\Delta=0.55\pm$0.11, and $\xi=0.12\pm$0.03.
Diffusive reacceleration is also accounted, although our propagation scenario favors models with no reacceleration.
The propagation calculations are made for all nuclear species in \GCR from $Z=1$ to $Z=26$.
All LIS fluxes are evaluated at the position of the Solar System, in cylindrical
coordinates $z_{\odot}\cong$0 of height and $r_{\odot}\cong$8.3\,kpc of galactocentric radius.
The calculations are described in details in our past works \citep{Tomassetti2018PHeVSTime,Tomassetti2019Numerical,Feng2016}. 
The key model parameters are constrained using LIS data from Voyager-1 \citep{Cummings2016},
high-energy measurements on primary nuclei (p-He-C-O) from \AMS \citep{Aguilar2015Proton,Aguilar2015Helium},
and secondary to primary ratios \citep{Aguilar2016BC,Aguilar2021Report}.
%
\begin{figure}[!t]
\centering
\includegraphics[width=0.46\textwidth]{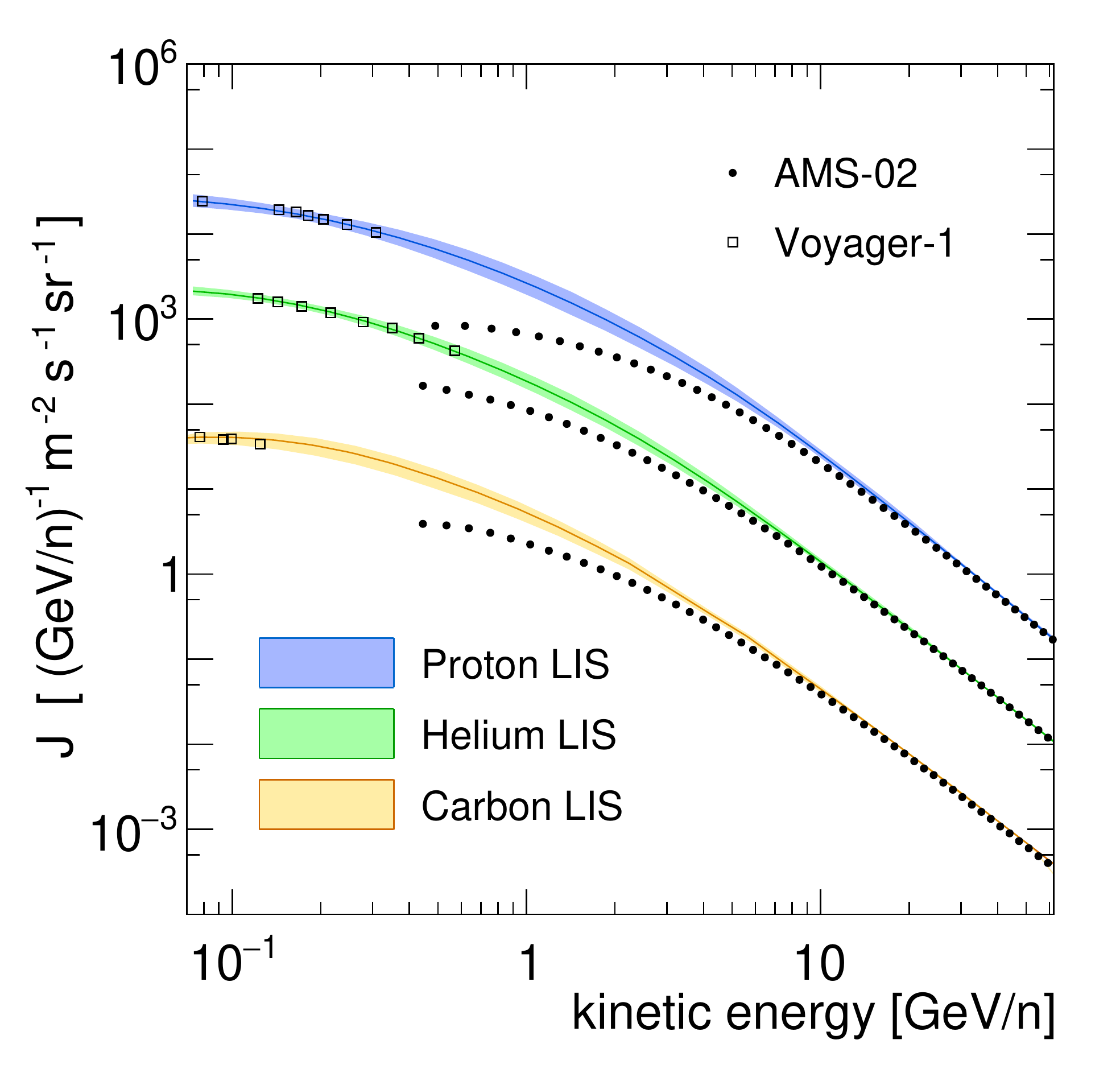}
\caption{\footnotesize{%
    LIS calculations for \GCR proton, helium, and carbon fluxes in comparison with direct measurements
    from \AMS{} in the ISS and Voyager-1 in interstellar space \citep{Cummings2016,Aguilar2015Proton,Aguilar2015Helium}.
}}
\label{Fig::ccLISProtonHeliumCarbon}
\end{figure}
%
The calculated LIS fluxes are illustrated in Fig.\,\ref{Fig::ccLISProtonHeliumCarbon} in comparison
with the data for the three most abundant \GCR species.

\subsection{Modeling the TOA fluxes} 
\label{Sec::TOAModel}                
%
%
We now present the calculations for the time-dependent modulation of \GCRs near Earth.
The solar modulation effect is determined by basic transport processes inside the heliosphere.
To compute the \GCR fluxes at a given epoch and in a given location in the heliosphere,
one has to solve the Parker's equation of \GCR transport. In its general form, the equation accounts for
diffusion in the small-scale irregularities of the interplanetary magnetic field,
gradient and curvature drift across its large-scale components, convection with the solar wind and adiabatic energy losses \citep{Potgieter2013}.
Here  employ the simple \emph{force field} (FF) solar modulation model. The FF model
is a simplified solution of the Parker's equation for a spherically symmetric wind
and an isotropic diffusion coefficient. The model provides a simple one-to-one correspondence between TOA 
and LIS fluxes in terms of a lower shift in kinetic energy of the \GCR density \citep{GleesonAxford1968,CaballeroLopez2004}.
For a \GCR nucleus of mass $M$, atomic number $Z$ and mass number $A$,
the TOA flux $J$ is related to its interstellar value $J^{\rm{IS}}$ by the relation:
\begin{equation}\label{Eq::GleesonAxford} 
  J(E) = \frac{E(E+ 2 m_{p})}{ (E+\Phi)(E+\Phi+2 m_{p}) } \times J^{\rm IS}(E + \Phi) \,,
\end{equation}
where $E\equiv{T/A}$ is the kinetic energy per nucleon, and $m_{p}\cong{M/A}$ is the nucleon mass.
The sum ${E}+\Phi$ is the kinetic energy per nucleon of \GCRs outside the heliosphere.
The parameter $\Phi\equiv (eZ/A)\phi$ represents the mean kinetic energy loss of \GCRs in the heliosphere,
and $\phi$ is the so-called \emph{modulation potential}.
The \GCR modulation of all \GCR nuclei is then described by a unique parameter, $\phi$, 
which has the dimension of an electric potential or a rigidity.
The modulation potential is related to physical quantities such as diffusion
coefficient $K$, solar wind speed $V$, and helispheric boundaries $r_{\rm hp}$ as:
\begin{equation}\label{Eq::PhiMeaning}
  \phi = \int_{r_{0}}^{r_{\rm hp}}  \frac{{V}(r)}{3 K(r)} dr \,,
\end{equation}
where $r_{0}=1$\,AU is our location.
With a quasi-stationary approach, the long-term variations of the \GCR fluxes can be expressed in terms of a
time-dependent modulation parameter $\phi=\phi(t)$. 
Several strategies have been developed for the reconstruction of the modulation level $\phi(t)$
at different epochs \citep{Maurin2015,Usoskin2011}. 
The FF model suffers from severe limitations \citep{CaballeroLopez2004,Tomassetti2017Universality}.
It does not capture physical mechanisms such as drift or anisotropic diffusion, 
that require more advanced modeling \citep{Boschini2017,Boschini2019,Potgieter2014,Fiandrini2021}.
Another problem is the applicability of the quasi-steady-state approximation \citep{Tomassetti2019Numerical}.
We emphasize, however, that for the purpose of this work the $\phi$ parameter is
regarded a mere proxy for the time variations of the \GCR modulation.
Thus, it is not regarded as a physical quantity representing the conditions of the heliospheric plasma.
As we will see, the convenience of the FF model is to express the time-dependent response of different detectors
into comparable time series of a unique parameter.

\subsection{Modeling the NM response}  
\label{Sec::NMRateModel}               

We now present a parametric description of the NM rates and their link with the time-dependent \GCR fluxes. 
The rate $\mathcal{R}_{\rm NM}^{d}$ of a NM detector $d$ at given epoch $t$ after correction for the barometric
pressure is given by \citep{Usoskin2011,Gil2015,Mishev2020}:  
\begin{equation}\label{Eq::NMRate}
  \mathcal{R}^{d}(t) = \sum_{j={\rm{\GCRs}}}\int_{0}^{\infty}  dE \cdot \mathcal{H}^{d}_{j}(E) \cdot \mathcal{Y}^{d}_{j}(E)\cdot J_{j}(t,E) \,.
\end{equation}
The sum is extended to all contributing \GCR species, where the solar modulated flux of the $j$-th species is $J_{j}(t,E)$.
In practice we consider only proton and helium. Heavier nuclei such as carbon or oxygen
contribute to a $\sim$\,1\,\% of the NM rate, and with a very similar temporal dependence.
The $\mathcal{Y}^{d}_{j}(E)$ function is the so-called yield function, measured in $m^{2}\,sr$.
It captures the energy-dependent response of a NM to $j$th-type \GCRs at unit intensity.
It includes the physics of hadronic showering in the atmosphere,
the detection efficiency of the instrument, its absolute normalization, its dependence on altitude \citep{Cheminet2013}.
In this work, to compute the NM yield, we implemented the parametric model of \citet{Maurin2015} (see Sec.\,4.2). 
The factor $\mathcal{H}^{d}$ in Eq.\,\ref{Eq::NMRate} is a \emph{transmission function}. It accounts for
the geomagnetic field modulation of \GCRs. We model it as a smoothed Heavyside function of rigidity \citep{SmartShea2005}:
\begin{equation}\label{Eq::GMTransmission}
  \mathcal{H}_{j}^{d}(E) = \frac{1}{ 1 + \left[ R_{j}(E)/R^{d}_{C} \right]^{-s} } \,, 
\end{equation}
where $R^{d}_{C}$ is the geomagnetic cutoff rigidity of a given NMs station and $s={12}$ sets the sharpness of the transition.
The transmission function is assumed constant in time, although the parameter $R^{d}_{C}$ is subjected to slow secular variations \citep{Cordaro2019}.
The transmission function is also particle independent when expressed as function of rigidity, $\mathcal{H}(R)$.
In Eq.\,\ref{Eq::GMTransmission}, the $j$-dependence arises from the relation between rigidity and kinetic energy per nucleon,
\begin{equation}\label{Eq::EknToRig}
  R_{j}=(A_{j}/Z_{j})\left[ E^{2} + 2m_{p}E \right]^{1/2}\,,
\end{equation}
which contains the mass/charge ratio of the $j$-th specie. 
We also note that, due to geomagnetic transmission, the energy integration of Eq.\,\ref{Eq::NMRate} contributes for $E\gtrsim\,E^{j}_{C}$,
\ie, above the energy corresponding to the local cutoff $R_{C}$:
\begin{equation}\label{Eq::RigToEkn}
E^{j}_{C} = \left[ R_{C}^{2} \left(Z_{j}/A_{j}\right)^{2} + m_{p}^{2} \right]^{1/2} - m_{p} \,.
\end{equation}
For this reason, other authors set a lower limit $E_{C}$ in the integration of the NM rate,
in place of using the $H$ function \citep{Gil2015,Mishev2020,Maurin2015}.
Once the NM response is fully specified, variations of the NM rates $\mathcal{R}_{\rm NM}^{d}(t)$
can be directly related to the time dependence of the TOA \GCR fluxes $J_{j}=J_{j}(t,E)$. 
Moreover, as seen in Sec.\,\ref{Sec::TOAModel}, the modulated fluxes can be related to a unique parameter $\phi$.

\section{Data Analysis and Results}  
\label{Sec::Results}                 

\begin{figure*}[!t]
\centering
\includegraphics[width=0.88\textwidth]{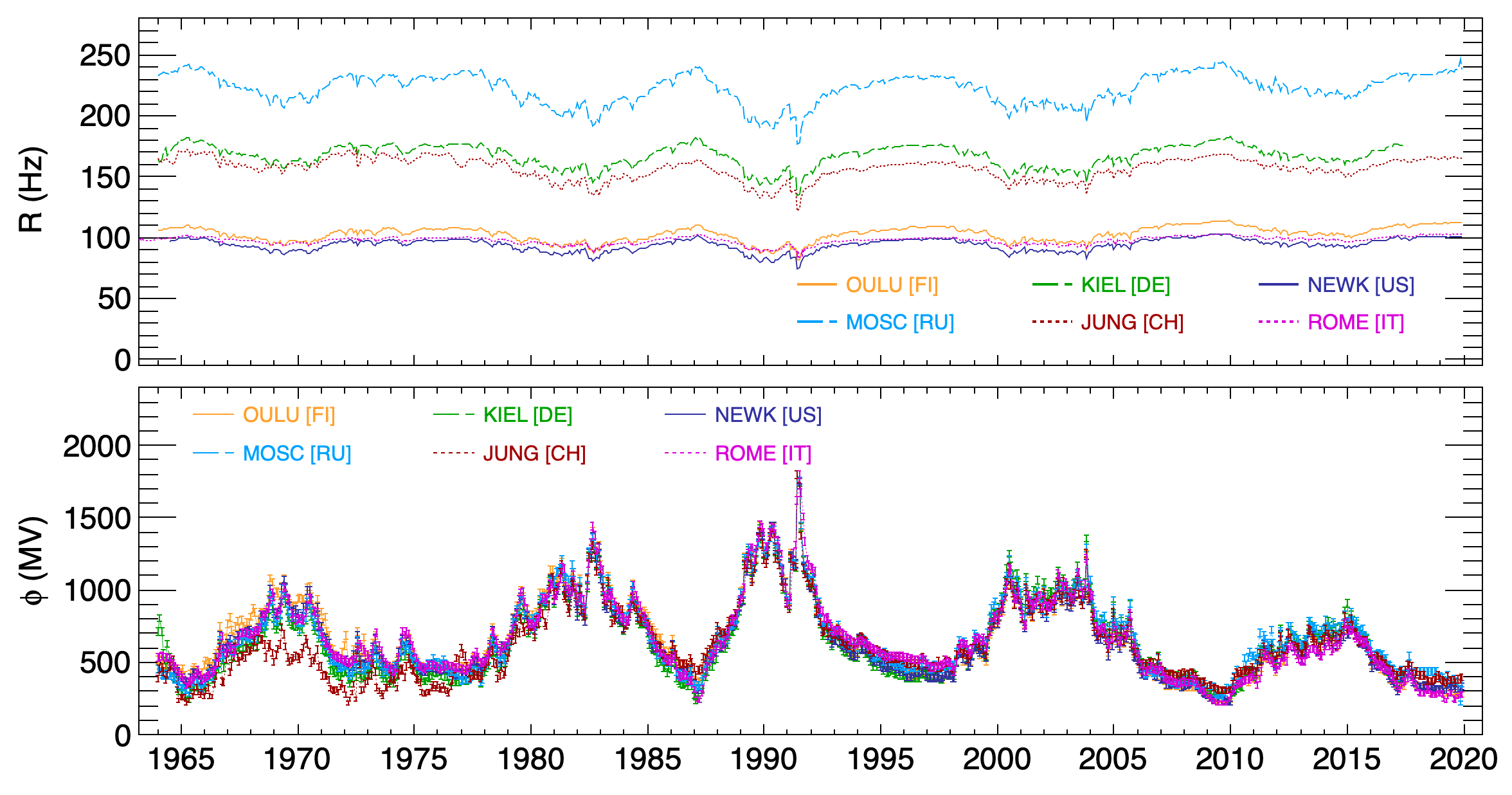} 
\caption{\footnotesize{%
    Temporal dependence of NM counting rates $\mathcal{R}^{d}(t)$ for six detector stations (top) and the
    corresponding reconstruction of the modulation potential $\phi^{d}(t)$ (bottom) on monthly basis from 1965 to 2020.
}}
\label{Fig::FromRateToPHi}
\end{figure*}

\subsection{Time series}  
\label{Sec::TimeSeries}   
%
%
\setlength{\tabcolsep}{0.034in} 
\begin{table*}[!t]
\begin{center}
\small
\begin{tabular}{ccccccc}
\tableline
\tableline
NM station & \href{http://www01.nmdb.eu/station/newk/}{NEWK} & \href{http://www01.nmdb.eu/station/oulu/}{OULU} &  \href{http://www01.nmdb.eu/station/kiel/}{KIEL} & \href{http://www01.nmdb.eu/station/jung}{JUNG} & \href{http://www01.nmdb.eu/station/mosc/}{MOSC} & \href{http://www01.nmdb.eu/station/rome}{ROME}  \\
\tableline
Detector type & 6-NM64 & 9-NM64 & 18-NM64 & 3-NM64 & 24-NM64 & 20-NM64\\
Location & Newark, US  &  Oulu, FI & Kiel, DE & Jungfraujoch, CH & Moscow, RU & Rome, IT\\
Coordinates & 39.68\,N 75.75\,W & 65.05\,N, 25.47\,E & 54.32\,N, 10.12\,E & 46.55\,N, 7.98\,E & 55.47\,N, 37.32\,E & 41.86\,N, 12.47\,E \\
Altitude & 50\,m & 15\,m & 54\,m & 3570\,m  & 200\,m  & 0\,m\\
Cutoff & 2400\,MV  & 810\,MV  & 2360\,MV & 4500\,MV & 2430\,MV & 6700\,MV \\
\tableline
\end{tabular}
\caption{Main characteristics of the NM stations used in this work (from \url{http://www.nmdb.eu} \citep{Vaisanen2021}). \label{Tab::NMStations}} 
\end{center}
\end{table*}

With the framework presented in the previous section, time-dependent measurements of different
experiments have been converted into comparable time series of modulation potentials $\phi(t)$.
An important dataset consists in direct \GCR measurements on monthly basis from spacecraft. 
It includes helium flux measurements from the MED instrument onboard the IMP-8 satellite (from 1973 to 1997),
and carbon flux measurements from the CRIS experiment onboard ACE (from 1997 to 2020).
The ACE/CRIS data cover seven energy intervals between 59 to 200 MeV/n of kinetic energy per nucleon,
corresponding to 0.7 to 1.3 GV of carbon rigidity.
The IMP-8/MED data covers the range 140 - 380 MeV/n, corresponding to about 1 - 1.6 GV of helium rigidity. 
To combine the two datasets, we use the highest energy interval of the ACE time series.
We obtain a unique time series of \GCR flux measurements covering 48 years with a mean rigidity value $R\cong$\,1.25 GV. 
For each month, the data are fitted with the modulated flux of Eq.\,\ref{Eq::GleesonAxford} where the parameter $\phi$ is left as free parameter.
The minimizing function is given by:
\begin{equation}
  \chi_{i}^{2}(\phi)=\sum_{k}\left[ J^{\rm mod}_{i}(E_{k},\phi) - \hat{J}_{i}(E_{k})\right]^{2}\sigma_{i,k}^{-2} \,,
\end{equation}
where $\hat{J}_{i}(E_{k})$ is the $i$-th measurement of \GCR flux, along with its uncertainty $\sigma_{i,k}$.
In spacecraft data, statistical uncertainties are of the order of $\sim$\,5-10\%, while systematic uncertainties are 2-3\,\%.
The flux $J^{\rm mod}_{i}$ is FF-modulated using the input LIS of that species and calculated at the energies $E_{k}$ of the data.
In practice, the two series from ACE and IMP-8 cover distinct epochs, so that we end up
with a unique ACE+IMP-8 series $\phi_{i}=\phi(t_{i})$.
Similar time series have been derived using counting rates from NM detectors.
In this work, we have considered six stations: Oulu, Kiel, Newark, Moscow, Jungfraujoch, and Rome.
The main properties of these NM stations are summarized in Table\,\ref{Tab::NMStations}.
The considered stations have a large time span, covering at least 55 years of data, and different rigidity cutoff ranging from 0.8 to 6.7 GV.
The rigidity cutoff of a NM station depends on its location in the local geomagnetic field.
It is calculated as the vertical Stoermer cutoff \citep{SmartShea2005}.
The data of each station consist in monthly averaged rates.
All rates are corrected for barometric pressure and detection efficiency \citep{Vaisanen2021}. 
For each dataset $d$, the corresponding time series of $\phi^{d}$ has been determined with a minimization procedure.
Given the measured rate for the $i$-th month $\hat{\mathcal{R}}_{d,i}$, the minimizing function is:
\begin{equation}
  \chi^{2}_{d,i}=\left[ \frac{\mathcal{R}^{d}_{i}(\phi) - \hat{\mathcal{R}}^{d}_{i}}{\sigma^{d}_{i}} \right]^{2} \,,
\end{equation}
where $\hat{\mathcal{R}}^{d}_{i}(\phi)$ is calculated in Eq.\,\ref{Eq::NMRate}
and its dependence on $\phi$ is contained in $J_{i}^{d}$, see Eq.\,\ref{Eq::GleesonAxford}. 
The $\sigma^{d}_{i}$ factors are the total uncertainties in the NM rates, that are of the order of $\sim$\,8-10\,\% \citep{Maurin2015}. 
From the NM rates, we obtain seven time series $\phi_{d}(t_{i})$ over the same time period that can be compared with each other.
The measured NM rates are shown in the top panel of Fig.\,\ref{Fig::FromRateToPHi}. 
The corresponding time series of modulation potential are shown in the bottom panel.
%
\setlength{\tabcolsep}{0.05in} 
\begin{table*}[!th]
\begin{center}
\small
\begin{tabular}{cccccccc}
\tableline
\tableline
\tableline
Parameters   &  IMP-8 + ACE  &  OULU  &  KIEL  &  NEWK  &  MOSC  &  JUNG  &  ROME  \\ 
\tableline
 $\tau_{M}$ (months) &  9.82 $\pm$ 0.42  &  7.76 $\pm$ 0.40  &  7.37 $\pm$ 0.45  &  7.36 $\pm$ 0.38  &  7.56 $\pm$ 0.41  &  7.99 $\pm$ 0.52  &  8.33 $\pm$ 0.41  \\ 
 $\tau_{A}$  (months) &  4.87 $\pm$ 0.55  &  5.22 $\pm$ 0.55  &  4.99 $\pm$ 0.60  &  5.00 $\pm$ 0.52  &  5.31 $\pm$ 0.56  &  5.64 $\pm$ 0.71  &  4.76 $\pm$ 0.56  \\ 
 $T_{0}$  (years) &  21.44 $\pm$ 0.73  &  21.20 $\pm$ 0.58  &  22.22 $\pm$ 0.85  &  21.04 $\pm$ 0.58  &  21.56 $\pm$ 0.63  &  20.99 $\pm$ 0.75  &  21.63 $\pm$ 0.67  \\ 
 $t_{P}$  (years) &  2.25 $\pm$ 0.51  &  1.50 $\pm$ 0.44  &  3.09 $\pm$ 0.64  &  1.42 $\pm$ 0.43  &  1.61 $\pm$ 0.44  &  1.32 $\pm$ 0.55  &  2.05 $\pm$ 0.52  \\ 
\tableline
\tableline
\end{tabular}
\caption{Summary of the fit results on individual datasets with the parameters of Eq.\,\ref{Eq::TimeLagVSTime}.\label{Tab::FittingResults}}
\end{center}
\end{table*}
%
All time series agree fairly well within the uncertainties.
Little discrepancies or glitches can be noted for a few stations.
They can be related to unaccounted inefficiencies, dead time, or other sources for unstable behavior \citep{Mishev2020}.
Nonetheless, the NMs operated with high stability, overall, during the long observation period of this work.

\begin{figure*}[!th]
\centering
\includegraphics[width=0.88\textwidth]{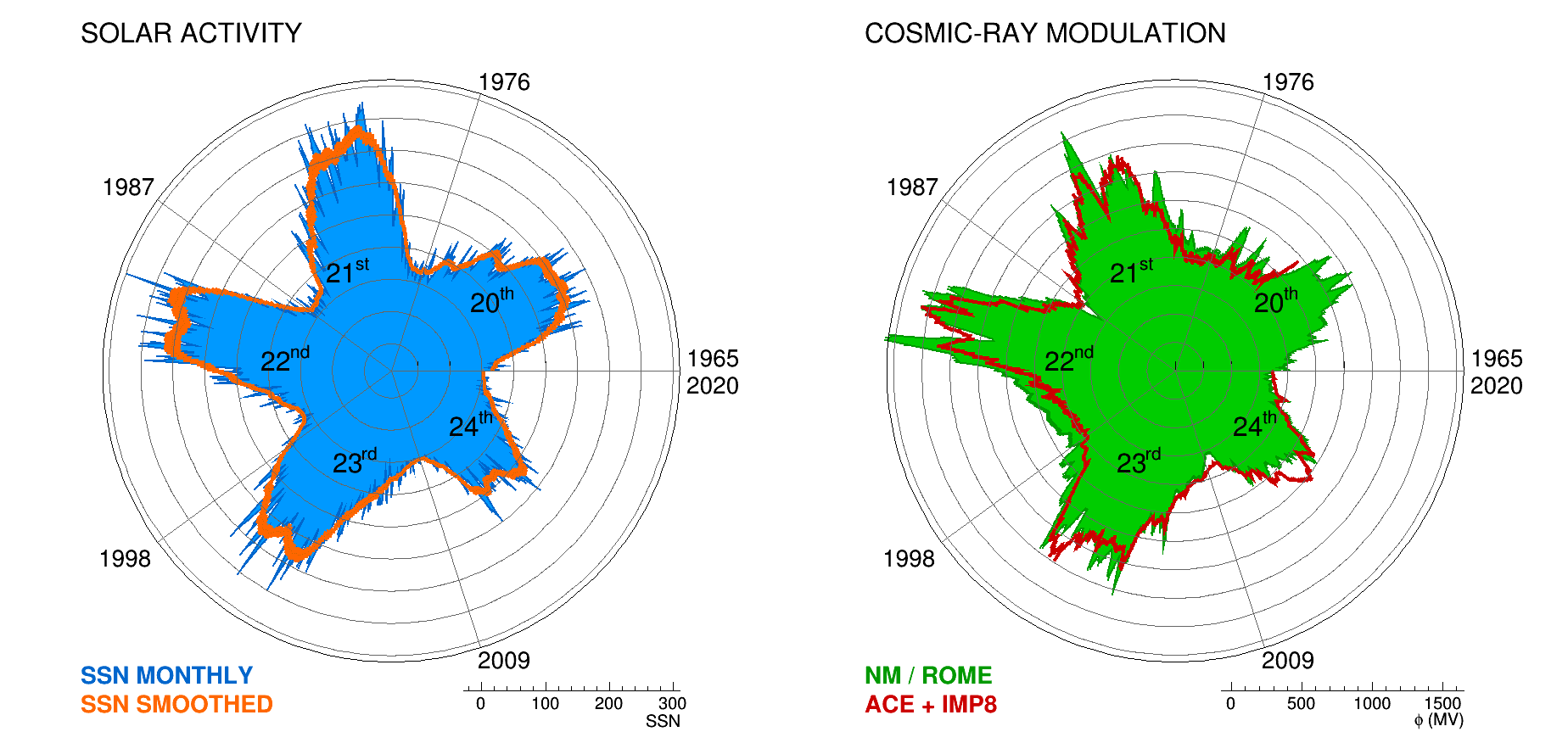}
\caption{\footnotesize%
  Temporal evolution of the monthly SSN (blue) and its smoothed value $S(t)$ (orange) since 1965 to 2020. Right: evolution of the monthly
  modulation potential $\phi$ reconstructed using NM data from the SVIRCO station in Rome, Italy (green) and using
  direct \GCR measurements in space made from IMP-8 (1973-1997) and ACE (1997-2020).
}
\label{Fig::ccPolarSSN}
\end{figure*}
%
As a proxy of solar activity, we consider the SSN data provided on monthly basis (calendar months) \citep{CletteLefevre2016}.
Thus, we define a continuous function $S(t)$ that interpolates the 13-month smoothed SSN series.
In the left panel of Fig.\,\ref{Fig::ccPolarSSN}, the time evolution of the monthly (blue) and smoothed (orange)
SSN is shown for five solar cycles.
The SSN data are compared with the reconstruction of the \GCR modulation potential $\phi$.
The modulation parameter is shown in the right panel for two datasets:
spacecraft data on \GCR fluxes from MED/IMP-8 and CRIS/ACE (red line), and NM rates from the Rome station (green area).  
In the figure, the starlike structure of the graphs reflects the quasiperiodicity of the solar cycle.
The $\phi$ time series from spacecraft data shows a similar structure of those obtained with NM rates.
Some little discrepancies are observed during solar maxima.
They can be ascribed to known features of the FF model in the low-energy region \citep{Tomassetti2017BCUnc,Gieseler2017}. 
Moreover, the two plots show a similar starlike structure. This reflects the known anticorrelation between SSN and \GCR flux.
However, here we observe a \emph{positive} correlation because we consider the modulation potential in place of the \GCR flux.

\subsection{Time lag analysis} 
\label{Sec::TimeLagAnalysis}   

\begin{figure}[!t]
\centering
\includegraphics[width=0.46\textwidth]{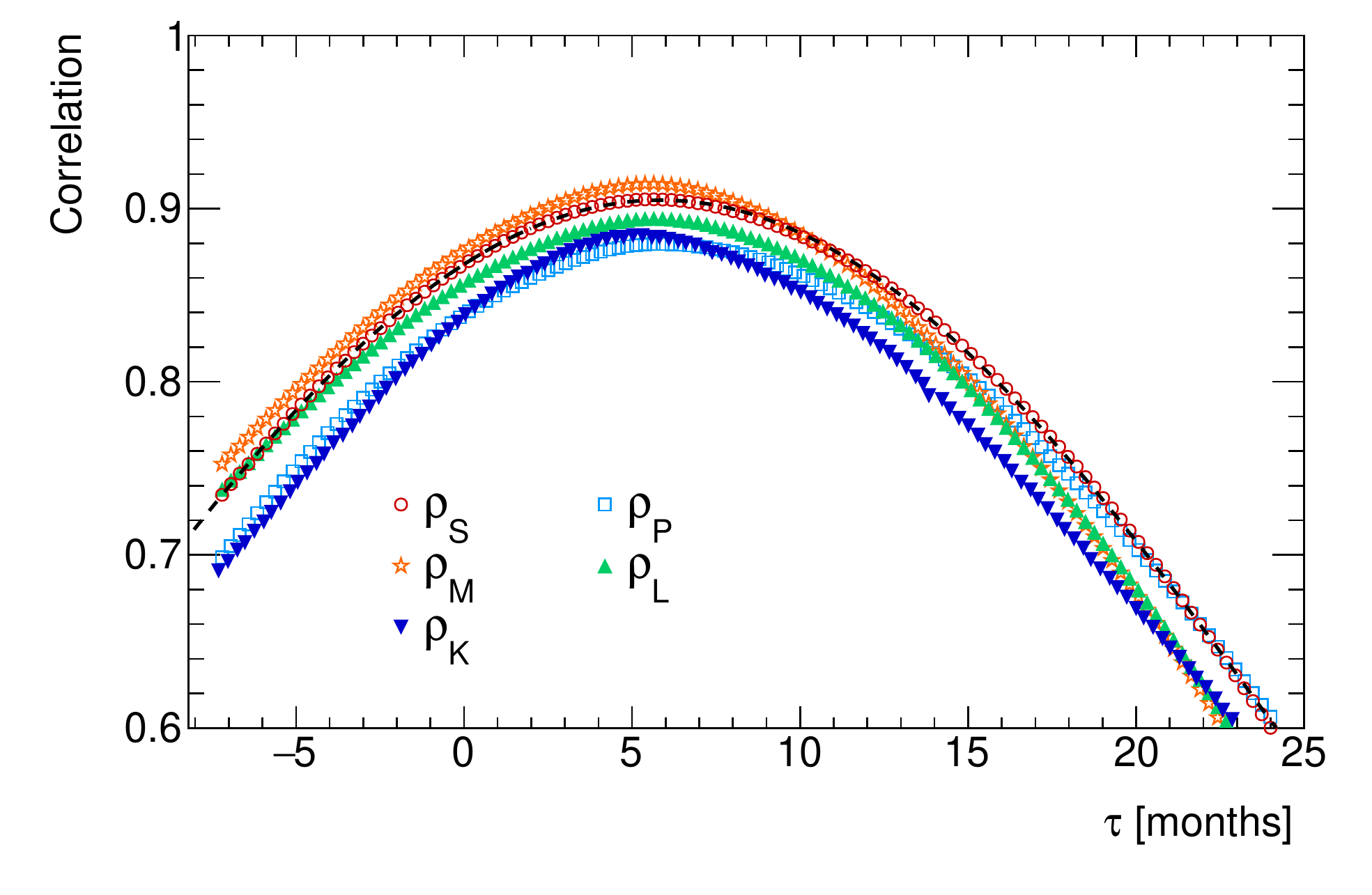}
\caption{\footnotesize{%
    Correlation coefficients evaluated as function of the assumed time-lag $\tau$ between SSN and \GCR flux.
    The graphs have been obtained with the Oulu NM dataset between 1986 and 1988. A Gaussian fit is shown
    for the curve obtained with the Spearmann coefficient.
}}
\label{Fig::ccCorrelationCoefficients}
\end{figure}

From the existence of a time lag $\tau$ between solar activity and \GCR flux modulation, we expect that the
parameter $\phi$ calculated at epoch $t$ should be maximally correlated with the SSN observed at a previous epoch $t-\tau$.
This can be noticed from the graphs in Fig.\,\ref{Fig::ccPolarSSN}. Here the \emph{modulation star} (right)
shows a small tilt toward counterclockwise direction in comparison with the \emph{sunspot star} (left).
Note that a rigid tilt would reflect and unique and constant time shift.
In the present work, we investigate whether and how the modulation lag \emph{evolves} over the solar cycle.
To determine the lag $\tau_{d}$ between the $d$th time series $\phi^{d}$ and the SSN,
we use a criterion based on the maximum correlation. 
For a set of observations on \GCR fluxes and SSN in a given time interval, our best estimate of the
lag is the parameter $\hat{\tau}_{d}$ for which the degree of correlation between  $\phi^{d}(t)$ and $S(t-\tau_{d})$ is maximum.
Note that as proxy for solar activity we use the smoothed SSN series.
This allows us to work with a continuous function $S(t-\tau)$.
To evaluate the degree of correlation with a given lag $\tau$, we make use of the
Spearman's rank-order correlation coefficient as default, $\rho^{d}_{S}=\rho^{d}_{S}(\tau)$. 
The procedure is based on a running window technique and it is outlined as follows.
First, we define a time grid $\{t_{k}\}_{k=1}^{N}$ of $N$ equidistant epochs (with constant pitch $p_{t}{\equiv}t_{k+1}-t_{k}$)
over the total observation period.
For each epoch $t_{k}$, we build a $\delta_{t}$-sized window $[t_{k}-\delta_{t}, t_{k}+\delta_{t}]$.
Thus, we analyze the $\phi$-SSN correlation inside that window.
More precisely, for every dataset $d$ we compute the correlation coefficient $\rho^{d}_{S}(\tau)$ between two sets of data:
all monthly evaluations of $\phi^{d}(t)$ where $t$ lies in the considered window,
and their corresponding SSN values $S(t-\tau)$ evaluated at the time $t-\tau$.
At this point, to estimate the lag, we repeat the procedure multiple times by varying the lag parameter.
In practice we make a tight scan between $\tau=-6$ and $\tau=24$ months. 
For all the considered epochs, the resulting function $\rho^{d}_{S}(\tau)$ appears to be remarkably Gaussian-shaped around a maximum value.
The maximum $\hat{\tau}_{d}$ is taken as the best estimate of the lag in the considered epoch $t$, for the dataset $d$.
Typical values are  $\hat{\tau}_{d}\sim$\,3-12 months.
The advantage of the Spearman's coefficient lies on its independence on any functional relationship between the two variables.
Along with Spearman, other correlation coefficients have been tested.
The dependence of various correlation coefficients on the lag parameter $\tau$
are shown in Fig.\,\ref{Fig::ccCorrelationCoefficients} for the Oulu dataset.
The following coefficients are shown in the figure:
the Pearson's product-moment correlation $\rho^{d}_{P}$, which is a measure of the \emph{linear} correlation between $\phi^{d}$ and SSN;
the Matisse $\rho^{d}_{M}$, which measures the linear correlation between $\phi^{d}$ and
the logarithm of SSN, as discussed in \citet{Tomassetti2017TimeLag};
the $\rho^{d}_{L}$ \emph{log-log} coefficient which measures the Pearson correlation of both logarithms of $\phi$ and SSN;
the Kendall rank correlation coefficient $\rho^{d}_{K}$;
finally, the Spearman's correlation coefficient $\rho_{S}(\tau)$ which is shown together with a Gaussian fit. 
From Fig.\,\ref{Fig::ccCorrelationCoefficients}, in can be seen that all the correlation methods agree
with a maximum correlation at $\hat{\tau}\approx{5-6}$\,months. 
The procedure is applied to the whole observation periods and for all the time series.
For each dataset $d$, we end up with a time series of lag values $\tau^{d}$.
%
%
%
%
\begin{figure*}[!th]
\centering
\includegraphics[width=0.88\textwidth]{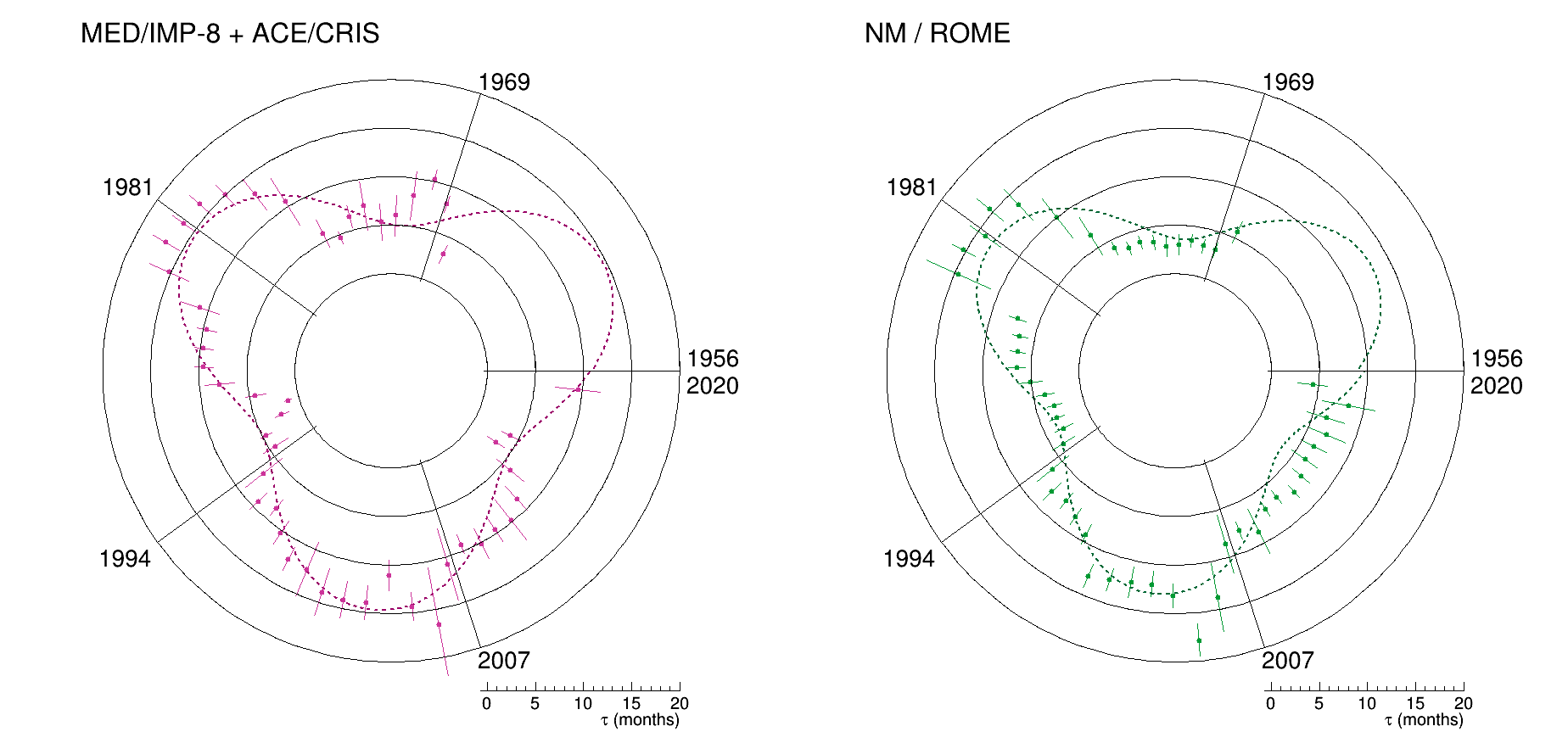}
\caption{\footnotesize%
    Temporal evolution of the \GCR modulation lag reconstructed on yearly basis over five solar cycles.
    The two reconstructions are done using \GCR data from ACE/IMP-8 (left) and from the NM station in Rome (right).
    The dashed lines are a sinusoidal fit of Eq.\,\ref{Eq::TimeLagVSTime}, showing a best-fit period $T\approx{22}$\,years.
}
\label{Fig::ccPolarTimeLag}
\end{figure*}
%
%
%
\begin{figure*}[!ht]
\centering
\includegraphics[width=0.88\textwidth]{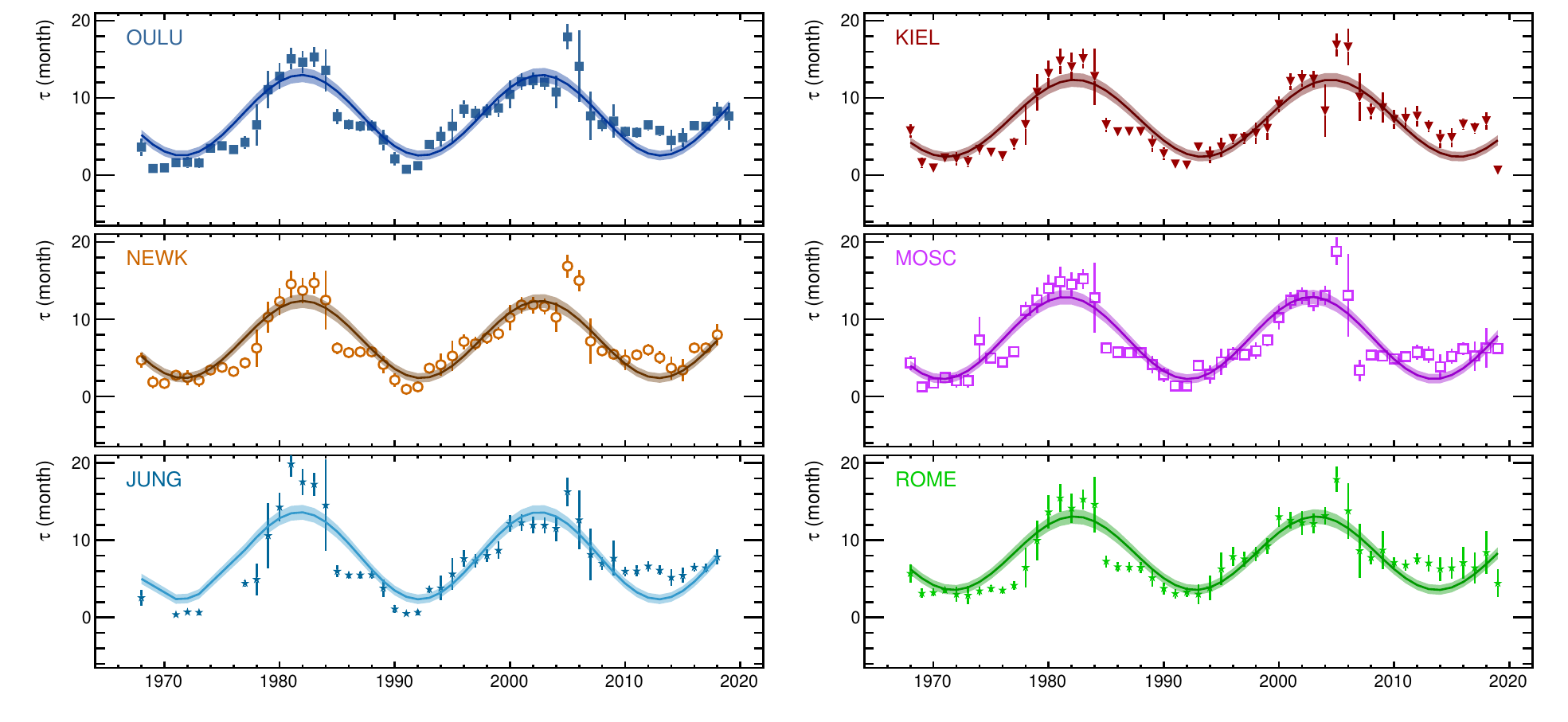}
\caption{\footnotesize%
    Temporal evolution of the \GCR modulation lag reconstructed on yearly basis over five solar cycles.
    The reconstructions are obtained from NM rates of the six stations in Table\,\ref{Tab::NMStations}.
    The solid lines show a fit to the data with Eq.\,\ref{Eq::TimeLagVSTime}, along with their with 68\,\% CL uncertainty bands.
}
\label{Fig::ccTimeLagVSTimeAllData}
\end{figure*}
%
%
%
\begin{figure}[!t]
\centering
\includegraphics[width=0.46\textwidth]{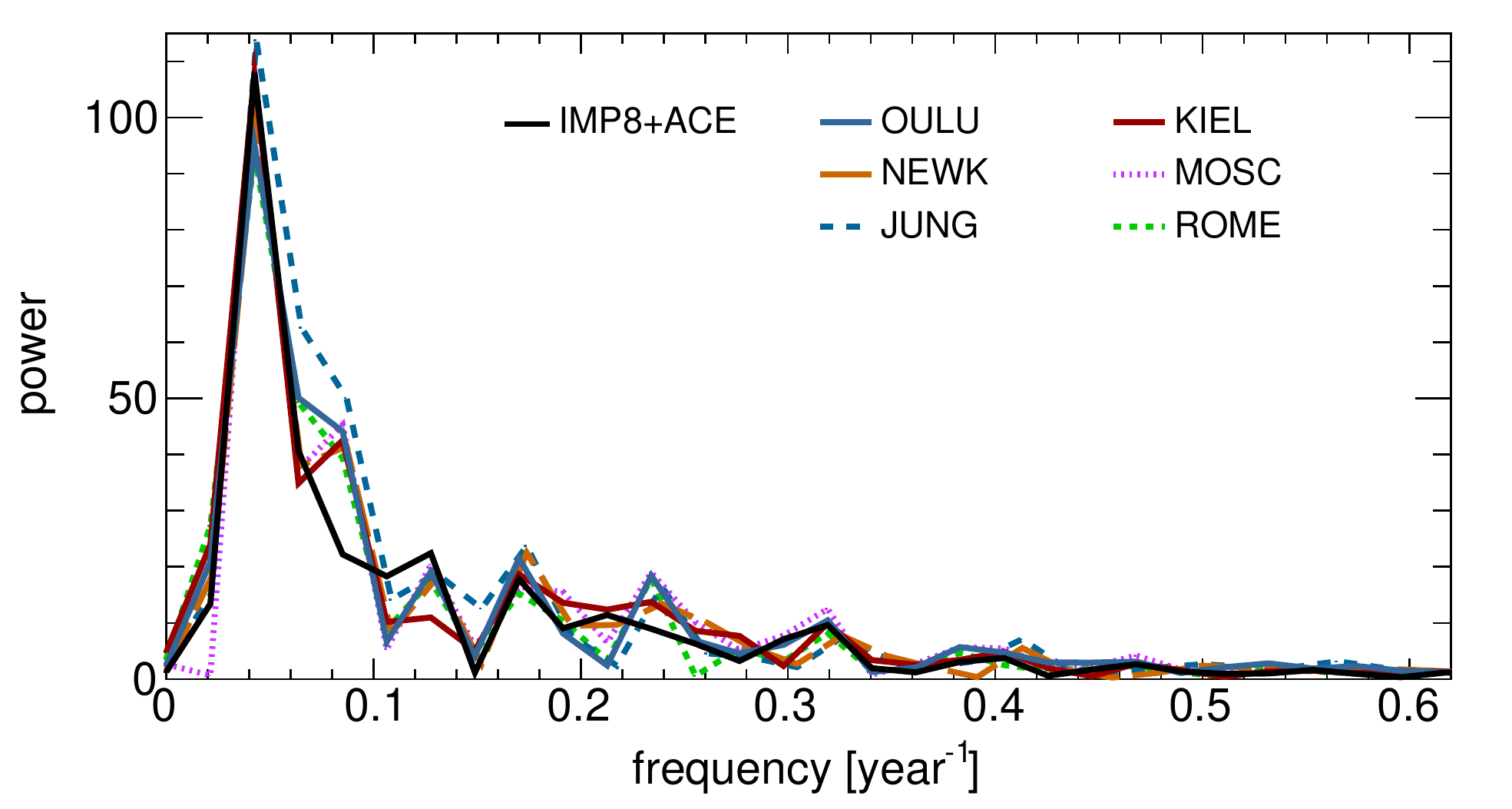}
\caption{\footnotesize{%
    Fast Fourier transform of the evolution of the modulation lag calculated using \GCR data from space (ACE+IMP8 combined) and
    from NM rates of several stations. All time series agree with the presence of a dominant frequency of 1/(22 year).}}
\label{Fig::ccTimeLagFFT}
\end{figure}
%
The results are found to be robust against the use of multiple datasets and different correlation coefficients.
They are also stable against the SSN smoothing parameters, the $\delta_{t}$ and $p_{t}$ parameters the time window. 
The impact of theoretical uncertainties in the LIS models, uncertainties on the smoothed SSN variance, 
in the correlation coefficients, in the determination of $\hat{\tau}$ was assessed. 
The reconstructed evolution of the lag is shown in Fig.\,\ref{Fig::ccPolarTimeLag}.
Here results are shown for \GCR data from space (left) and from the Rome NM station (right), over the last five solar cycles.
It can be seen that both time series show a remarkable periodicity.
In the figure, the data are fitted with a sinusoidal function (dashed line):
\begin{equation}\label{Eq::TimeLagVSTime}
\tau^{d}(t) = \tau_{M}^{d} + \tau_{A}^{d} \cdot \cos\left[ \frac{2\pi}{T^{d}_{0}}\left( t - t^{d}_{P}\right)\right] \,,
\end{equation}
where $\tau_{A}$ is the maximum amplitude of its variation, $T_{0}$ is the oscillation period,
and $t_{P}$ is the phase. The function oscillates around $\tau_{M}^{d}$, its average value.
Both datasets of Fig.\,\ref{Fig::ccPolarTimeLag} give a best-fit period $T_{0}=21.5{\pm}0.8$\,years.
The reconstructed lag from all NM stations is also given in Fig.\,\ref{Fig::ccTimeLagVSTimeAllData}.
The fits results with all the free parameters of Eq.\,\ref{Eq::TimeLagVSTime} are summarized in Table\,\ref{Tab::FittingResults}.
It can be seen that the different NM stations lead to consistent results.
We also note that some points seem to deviate from the periodical model, in particular around the minima/maxima.
Here the behavior looks a bit sharper than that of a simple sinusoidal.
In our opinion, some of these features may be related to irregularities in the solar cycle, while the overall trend
may be different from a perfect periodical and sinusoidal. 
Clearly, the sinusoidal model is a simplification of the actual trend, which could probably be described by better models.
Nonetheless, it captures well the 22-year periodicity of the modulation lag.
A further evidence for this periodicity is provided in Fig.\,\ref{Fig::ccTimeLagFFT}.
Here the Fourier transform of the modulation lag is plotted
for all time-series (NM and spacecraft data). 
The figure shows the 22-year periodicity peak as a dominant periodicity of the time-lag variation.
As an independent cross-check, we applied a different fit strategy. The parameters of Eq.\,\ref{Eq::TimeLagVSTime}
were determined by maximizing directly the correlation between $\phi^{d}(t)$ and $S(t-\tau^{d}(t))$.
In this approach, the smoothed SSN function $S$ depends on the $\tau^{d}(t)$ function, which is 
expressed as analytical function of the parameters $\Theta\equiv \{\tau_{M}, \tau_{A}, T_{0}, t_{P}\}$, from Eq.\,\ref{Eq::TimeLagVSTime}.
Thus, for any set of lag parameters $\Theta$, one can compute the correlation coefficient  $\rho^{d}_{S}=\rho^{d}_{S}(\Theta)$.
We define the minimizing function as $\tilde{\chi}^{d}(\Theta) \equiv 1 - |\rho^{d}_{S}(\Theta)|^{2}$.
The best-fit estimate of the lag parameters was obtained by the minimization of this function.
For all minimizations, we use \textsf{C++} routines of \texttt{MINUIT} implemented in the \texttt{ROOT} package.
This procedure has the advantage of providing directly the lag parameters.
There is no need of splitting the time series into subintervals and to reconstruct the whole lag evolution.
On the other hand, the results are no longer model independent.
They rely on the assumed functional dependence for the evolution of the lag (\ie, Eq.\,\ref{Eq::TimeLagVSTime}).
Nonetheless the two independent cross-checks gave consistent results. 

\subsection{Rigidity dependence} 
\label{Sec::RigidityDependence}  

The use of several datasets allows us to inspect the energy or rigidity dependence of the parameters.
In the following we make use of rigidity $R$.
The time series from spacecraft data is evaluated at at $R=$\,1.25\,GV.
The other time series from NM are unresolved on rigidity. Nonetheless, from Eq.\,\ref{Eq::NMRate},
one can compute the average rigidity of \GCRs producing the NM rates, which is of the order of dozens GV.
We then define the \emph{mean rigidity} $\langle{R^{d}}\rangle$, for the $d$-th dataset, as the time-averaged expectation value of $R$.
The average is calculated on \GCR energy spectrum that produces the observed rates. We write:
\begin{equation}\label{Eq::RigEff}
  \langle{R}^{d}\rangle \equiv \frac{1}{N^{d}} {\sum_{j}\int_{T} dt \int_{0}^{\infty}  R_{j}  \mathcal{H}^{d}_{j}  \mathcal{Y}^{d}_{j} J_{j} dE } \,,
\end{equation}
where all the quantities in the integrand function, including $R_{j}$, are expressed as function of kinetic energy per nucleon $E$. 
The $N^{d}$ factor represents the normalization and it is given by:
\begin{equation}\label{Eq::NormFactor}
  N^{d} \equiv \sum_{j}\int_{T}dt\int_{0}^{\infty}  \mathcal{H}^{d}_{j}  \mathcal{Y}^{d}_{j} J_{j} dE \,.
\end{equation}
Thus, the calculation of $\langle{R^{d}}\rangle$ involves the \GCR spectra, the NM yield function, the transmission function.
In particular, it depends on the geomagnetic rigidity cutoff $R^{d}_{C}$, which varies from NM to NM.
From Table\,\ref{Tab::NMStations}, the cutoff range from $0.8$\,GV (Oulu) to $6.7$\,GV (Rome).
Using NMs at different cutoff, one may study the rigidity dependence of NM rates \citep{Nymmik2000}. 
In fact the integral in Eq.\,\ref{Eq::RigEff} is suppressed at $R{\lesssim}R_{C}$, thus $\langle{R^{d}}\rangle$ increases with $R_{C}$.
However, this dependence is mitigated by the NM yield function $Y^{d}(R)$, which decreases rapidly with decreasing rigidity.
As a result, we found that all NMs give similar values for the man rigidity, between $\sim$\,25 and 35\,GV.
%
\begin{figure}[!t]
\centering
\includegraphics[width=0.46\textwidth]{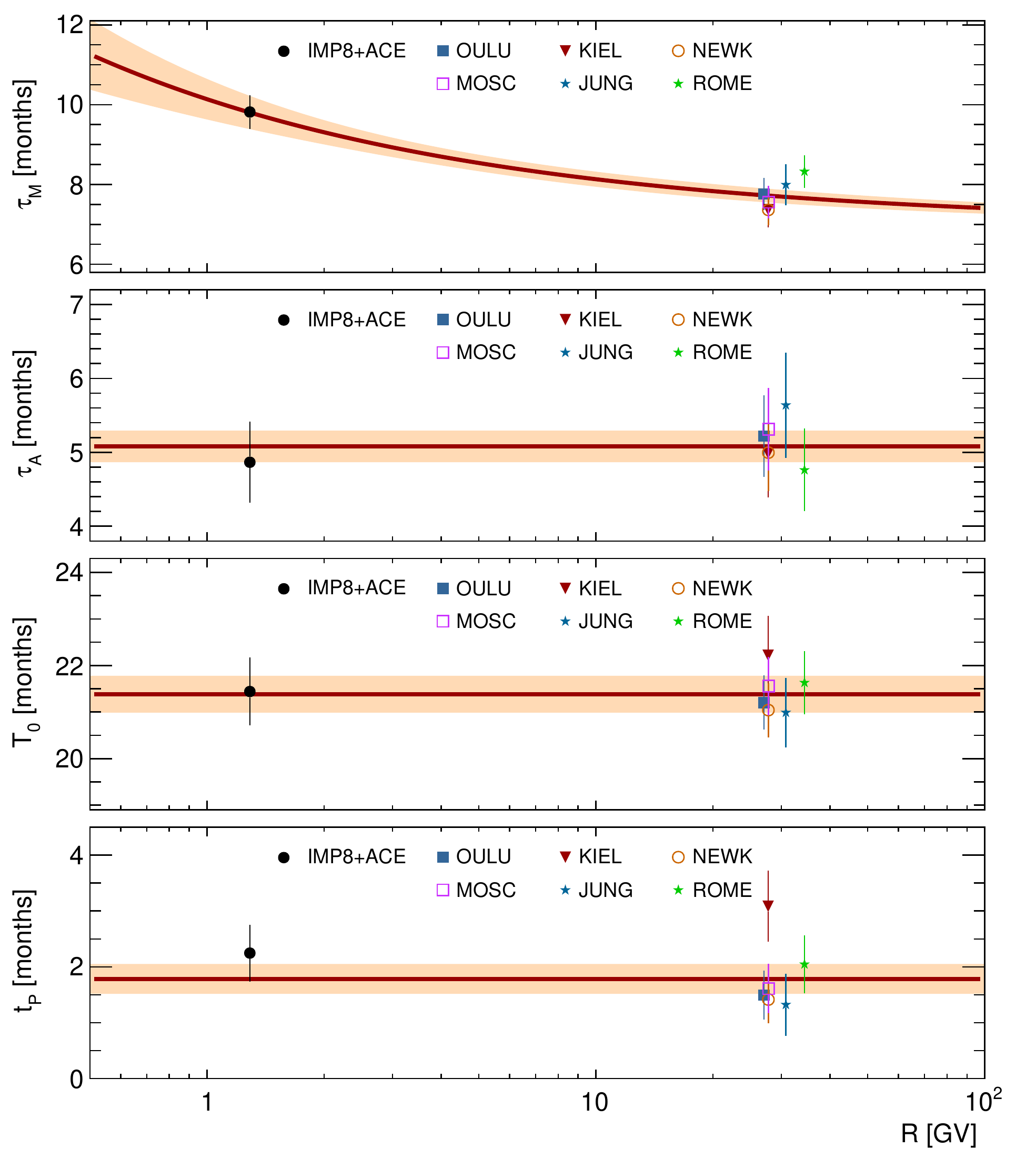}
\caption{\footnotesize{%
    Best-fit values of the lag parameters of Eq.\,\ref{Eq::TimeLagVSTime} for all time-series plotted against
    the mean rigidity $\langle{R}\rangle$. The solid lines show the global fit according to
    Eq.\,\ref{Eq::GlobalLagParameterization}, along with the 68\,\% CL bands. 
}}
\label{Fig::ccParametersVSRigidity}
\end{figure}
%
%
In Fig.\,\ref{Fig::ccParametersVSRigidity}, the best-fit parameters of Eq.\,\ref{Eq::TimeLagVSTime} are
plotted as function of $\langle{R^{d}}\rangle$.
From the figure, it can be seen that all NM data group together at $\langle{R^{d}}\rangle\sim$\,30\,GV.
As noted in Ref.\,\citep{RossChaplin2019}, from NM data alone it is difficult to determine any rigidity dependence.
Noticeably, all the datasets agree with a $T_{0}$ parameter fairly consistent with the period of a magnetic polarity cycle, 22 years. 
Similarly, phase and amplitude give consistent results, showing no indications for rigidity dependence.
In the figure, these parameters are fitted with constant functions.
On the other hand, the average lag value $\tau_{M}$ shows a remarkable \emph{decrease} with increasing \GCR rigidity.
From the fits, the low-$R$ data give an average lag $\tau_{M}$ of nearly ten months, while all NM data agree with a mean lag of about eight months.
To fit its rigidity dependence, we used a function of the type $\tau_{M}(R)=\tau^{0}_{\rm{Min}} + \tau_{M}^{0}(R/{\rm{GV}})^{-\alpha}$,
\ie, a power-law in rigidity $R$ with index $\alpha$ plus a constant offset $\tau^{0}_{\rm{Min}}$.
As we discuss in the next section, the choice of the power-law function is based on considerations on the rigidity dependence
of the \GCR diffusion timescale. The constant offset must be set to keep $\tau\geq{0}$ in the high-$R$ limit.

\subsection{A global formula for the modulation lag} 
\label{Sect::GlobalFormula}                          

\begin{figure*}[!t]
\centering
\includegraphics[width=0.88\textwidth]{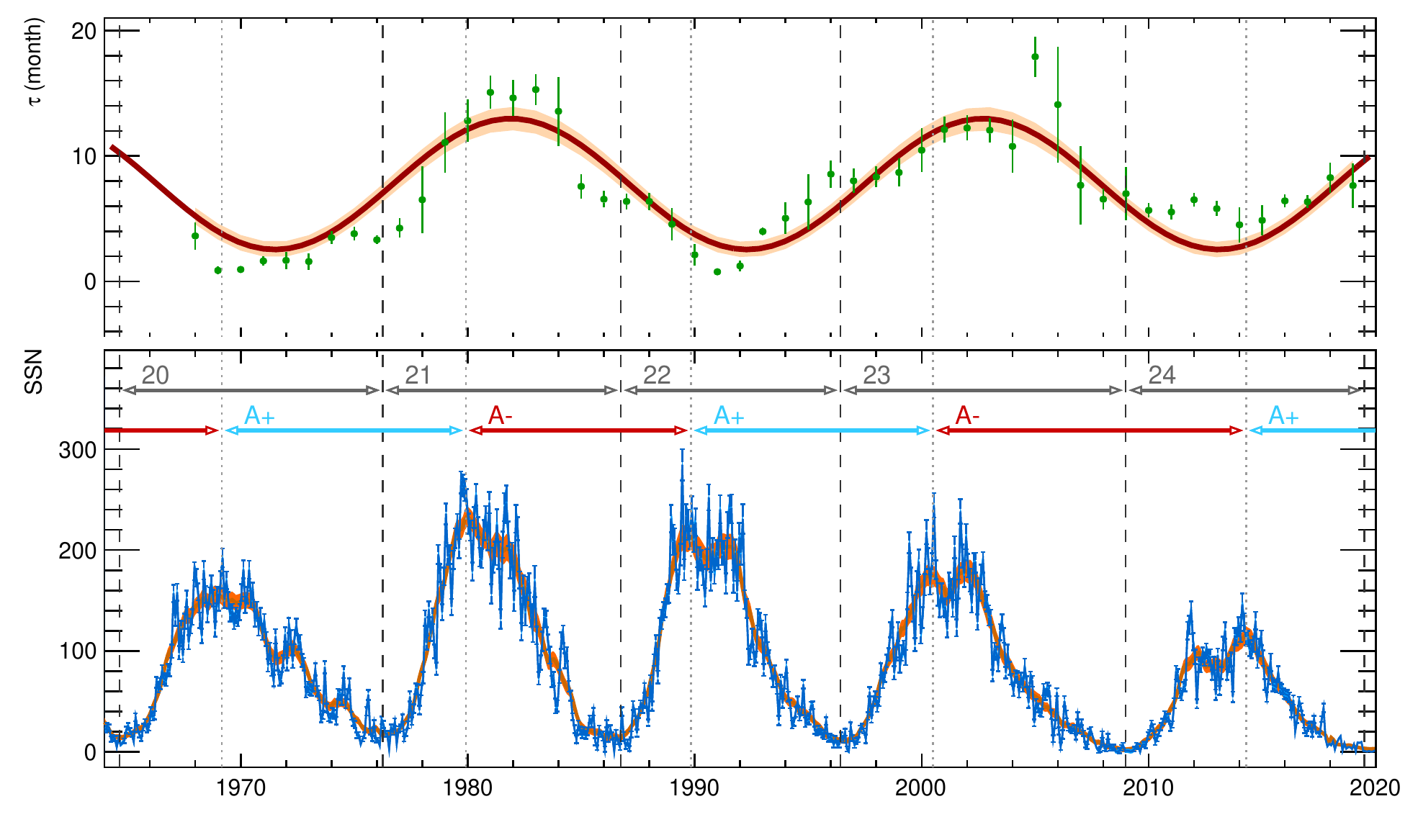}
\caption{\footnotesize{%
    Temporal evolution of the modulation lag determined using NM data from the Oulu station (top),
    in close comparison with the evolution of the SSN (bottom) over the last five solar cycles,
    The solid line shows the sinusoidal fit of Eq.\,\ref{Eq::TimeLagVSTime} and its associated uncertainties.
    The dashed and dotted lines indicate solar cycle minima and maxima, respectively, defining cycle numbers and their magnetic polarities.
}}
\label{Fig::ccTimeLagVSTime}
\end{figure*}

Based on the considerations made in the previous paragraph, we end up with the following formula: 
\begin{equation}\label{Eq::GlobalLagParameterization}
\tau = \tau^{0}_{\rm{Min}} + \tau_{M}^{0}\left(\frac{R}{\rm{GV}}\right)^{-\alpha} +  \hat{q}\tau_{A} \cos\left[ \frac{2\pi}{T_{0}}\left( t - t_{P}\right)\right]
\end{equation}
The equation describes the rigidity and temporal evolution of the modulation lag over the solar cycle.
The best-fit parameter values are the following:
$\tau_{M}^{0}=\,3.1\pm\,0.5$\,months,
$\alpha=0.5\pm\,0.09$,
$T_{0}=21.4\pm\,0.5$\,years, and
$t_{P}=1.8\pm\,0.2$\,months.
The equation is shown in Fig.\,\ref{Fig::ccParametersVSRigidity} (thick red line) along with its ``one $\sigma$''
uncertainty (yellow band) associated with the fit. 
In Eq.\,\ref{Eq::GlobalLagParameterization}, the factor $\hat{q}\equiv{q}/{|q|}$ is the charge sign of the \GCR particles.
The analysis presented here is based on positively charged particles, so that $\hat{q}=1$.
The charge-sign dependence of the lag is linked to the dependence of the modulation equations on the $\hat{q}\cdot{A}$ product
between \GCR charge-sign and solar magnetic polarity.
This dependence could be tested directly using time-resolved data on cosmic antiparticles such as,
in particular, \AMS measurements on \GCR antiprotons.
The formula can also be expressed in terms of cycle fraction $x\equiv{t/T_{0}}$, in place of the time variable,
which makes it useful to be applied or tested to other/future solar cycles. 
The equation is valid only for \GCR located near-Earth (at helioradius $r_{0}=$\,1\,AU).
Other regions of the heliosphere may have different lag parameters.  
In spite of its simplicity, this equation represents a generalization of the empirical delay factors used in many time-dependent
modulation models \citep{Slaba2020,Kuznetsov2017,Matthia2013,Tomassetti2017TimeLag}, which are often kept as constant.
The lag evolution over the different phases of the solar cycle is also illustrated in Fig.\,\ref{Fig::ccTimeLagVSTime}
and discussed in the following.
%

\section{Discussion}     
\label{Sec::Discussion} 

\subsection{Interpretation of the results} 
\label{Sec::Interpretation}                

The modulation lag is often interpreted in terms of timescales of the changing conditions of the heliosphere.
The process is regulated by the propagation speed $V$ of the solar wind, which fills a modulation region of size $L\sim$\,100\,AU.
The naive estimate suggests a modulation lag of the order of one year, $\tau\sim{L/V}$.
However, the present study suggests a more complex physical picture where also drift,
latitudinal dependence of the wind, and \GCR diffusion play an active role in the formation of the lag. 
A similar picture is also emerging from recent works \citep{Shen2020TimeLag,Wang2022TimeLag,Koldobskiy2022TimeLag},
as we discuss in the next paragraph.
Because of gradient or curvature drifts across the interplanetary magnetic field, \GCRs are guided
to follow preferential trajectories.
Hence, near-Earth observations of \GCRs probe only selected regions of the heliosphere. 
The drift speed of \GCRs, however, depend on the product $\hat{q}\hat{A}$.
During $\hat{A}^{+}$ epochs, cosmic protons reach the inner heliosphere through high-latitude regions.
During $\hat{A}^{-}$ epochs, they gather near the equatorial plane and across the heliospheric current sheet.
This phenomenon is coupled with the colatitudinal dependence of the solar wind.
The wind speed $V(\theta)$ ranges from $\sim\,$400\,km\,s$^{-1}$ near equators ($\theta{\sim}90^{\circ}$, slow wind region)
to $\gtrsim$\,800\,km\,s$^{-1}$ in the polar regions ($\theta{\sim}0^{\circ},180^{\circ}$, fast wind region) \citep{Potgieter2013}.
As a consequence, we expect larger lags when analyzing \GCR data in $\hat{q}\hat{A}<0$ conditions,
as we are probing the slow-wind region. 
In contrast, shorter lags are expected from data of $\hat{q}\hat{A}>0$, that probe the fast-wind regions of the heliosphere.
Moreover, we expect that matter and antimatter particles interchange their role with the change of polarity.
The propagation of \GCR protons (or positrons) under $\hat{A}^{+}$ periods should be similar to that of
\GCR antiprotons (or electrons) during $\hat{A}^{-}$ periods and viceversa.
This circumstance is expressed by the factor $\hat{q}$ in Eq.\,\ref{Eq::GlobalLagParameterization}
and could be resolutely tested using time-resolved data on \GCR antimatter.
The connection between the modulation lag and the 22-year polarity cycle is also illustrated in Fig.\,\ref{Fig::ccTimeLagVSTime}.
In the figure, along with the solar cycles, the $\hat{A}^{+}/\hat{A}^{-}$ magnetic polarities are shown.

In addition, the observed rigidity dependence suggests that the propagation times of \GCRs in
the heliosphere play a crucial role in the formation of the lag.
Here we refer to the time spent by \GCRs in the heliosphere before reaching Earth.
Their propagation time is dominated by parallel diffusion with a coefficient of the type $K\propto{R^\alpha}$,
where the index $\alpha$ is linked to the spectrum of magnetic turbulence \citep{Fiandrini2021}. 
Hence, the \GCR propagation time must decrease with rigidity as $T_{\rm{diff}}\sim{L^{2}}/3K(R){\propto}R^{-\alpha}$.
As a result, low-$R$ particles bring more ``delayed'' information in comparison with high-$R$ particles.
In particular, the diffusion time of 1\,GV particles is of the order of one month \citep{Gallagher1975,Strauss2011}. 
Such a rigidity-dependent lag, however, adds up to a rigidity-independent lag which is of 
the order of 6-8\,months.
After considering both terms, we end up with the final form of Eq.\,\ref{Eq::GlobalLagParameterization}.
Intriguingly, this result may offer a new way to measure the rigidity dependence of \GCR diffusion in heliosphere.
The parameter $\alpha$ of the present work agrees with other analysis based on \GCR propagation models \citep{Fiandrini2021},

\subsection{Comparison with other works}  
\label{Sec::Comparisons}                 

The study of the time lag in the long-term \GCR modulation was conducted in a number of recent works. 
Different techniques of data processing or analysis approaches were used \citep{Iskra2019,RossChaplin2019,Shen2020TimeLag}.
Several studies on the long-term \GCR modulation have pointed to an odd/even effect
for the lag. The lag observed during odd solar cycles appears to be longer than that in even ones.
These observations suggest the association of this feature with charge-sign dependent drift \citep{Usoskin1997,Usoskin2001}.
In this work, we determined the temporal evolution of the modulation lag over five solar cycles.
This shows its quasiperiodical behavior in connection with the 22-year solar cycle of magnetic polarity.
The role of drift is essential for the interpretation given in Sec.\,\ref{Sec::Interpretation}.
Two interesting papers appeared while this work was in preparation \citep{Wang2022TimeLag,Koldobskiy2022TimeLag}.
In \citet{Koldobskiy2022TimeLag}, the variability of the \GCR modulation lag is investigated by means
of a time-frequency wavelet analysis. A complete compilation of the past results is also presented,
showing the modulation lags reported at different epochs/cycles. Their results agree well with ours.
In \citet{Wang2022TimeLag}, it is suggested that the time lag between \GCR flux and SSN
originates already in the open magnetic flux on the Sun.
The lag between SSN and the generation of the open magnetic flux was also analyzed. 
The authors found a cycle-dependent lag that might explain the odd/even effect.
They also note that the role of \GCR transport should make a contribution.
In Sec.\,\ref{Sec::RigidityDependence} and Sec.\,\ref{Sec::Interpretation}, we also came to the conclusion
that the lag originates from different contributions. However, the delayed open flux scenario was not considered in our work.
In Sec.\,\ref{Sec::RigidityDependence}, to determine the rigidity dependence of the lag,
we have defined the mean rigidity $\langle{R}\rangle$ of \GCRs that forms the NM rates.
The notion of the effective \GCR rigidity (or energy) $R_{\rm{eff}}$ for NMs is discussed in many
works \citep{Alanko2003EffEnergy,Gil2017EffEnergy,Asvestari2017EffEnergy,Koldobskiy2018EffEnergy}.  
In \citet{Gil2017EffEnergy}, it is defined as the \GCR rigidity
at which protons have the same relative variability as the counting rate of the detector.
In \citet{Alanko2003EffEnergy}, 
$R_{\rm{eff}}$ is defined such that the NM rate is proportional to the flux of all \GCRs with rigidity $R$ \emph{above} this value.
In this paper, our definition of $\langle{R}\rangle$ is aimed at capturing the \emph{average} rigidity value
of the primary \GCRs that produce the NM signal.
Comparing the two definitions, one expects that $\langle{R}\rangle$ is some factor larger than $R_{\rm{eff}}$.
The rigidity dependence of the modulation lag is investigated first in \citet{Nymmik2000},
where the simple cutoff value $R_{C}$ was used as reference rigidity. 
We find, in agreement with this work, the general trend that odd solar cycles show longer lags than even cycles.
We did not find, however, any odd/even effect in the rigidity dependence of the lag.
Moreover, in agreement with \citet{RossChaplin2019}, we found that the NM data alone
do not lead to a clear determination of rigidity dependence.
It is also interesting to compare our work with the results of \citet{Shen2020TimeLag}.
In their work, the energy dependence of the lag was derived using only spacecraft data. Using IMP-8 proton data between
1980 and 1999, they found a decrease for $\tau(E)$ between about 50\,MeV and nearly 400\,MeV of kinetic energy.
These energies fall just below the range considered in our work.
In the highest region, corresponding to about 1\,GV of \GCR proton rigidity,
they reported a modulation lag of $\tau\sim$\,10\,months for the $\hat{A}^{-}$ dataset (epoch 1980-1989)
and  $\tau\sim$\,6\,months for the $\hat{A}^{+}$ dataset (epoch 1991-1999). Their results are in agreement
with our findings for the same considered periods.
However their analysis is only limited to the descending phases of the solar cycle. 
In \citet{Shen2020TimeLag}, the rigidity dependence of the modulation lag is described by a simple power-law function.
Their interpretation of such a dependence is in full agreement with ours.
However, they derived two distinct slopes for the two datasets of opposite magnetic polarity,
suggesting that the lag decreases more rapidly during $\hat{A}^{+}$ epochs.
The authors argue that a difference in the two slopes may be caused by different rates of \GCR energy losses in the two magnetic polarities.
Although such an interpretation appears reasonable to us, it should also be noted that the two trends are very similar each other. 
Within the uncertainties estimated in our analysis, the discrimination of the of the two slopes is not possible.
Regarding Eq.\,\ref{Eq::GlobalLagParameterization}, a comparison with past works can be done.
In models based on the transport equation, the time lag is usually treated as an input
parameter $\tau$ to be determined with the data \citep{Matthia2013,ZhaoQin2013,Zhu2018}.
In general, many models rely on the simplifying assumption of a constant lag \citep{Kuznetsov2017,ZhaoQin2013,Adams2011,Tomassetti2017TimeLag}.
In some recent works, the lag parameter is assumed to take different values for odd/even solar cycles,
\ie, it is modeled as a 22-year periodic step function \citep{Slaba2020,Kuznetsov2017}.
In the BOM model \citep{Slaba2020,ONeill2010}, the temporal evolution of the \GCR flux is calculated within the FF approximation.
The time dependence of the modulation parameter $\Phi(t)$ is calibrated against the SSN using a cycle-dependent delayed relationship.
In practice, a two-value lag is used as input parameter, depending on the odd/even cycle number.
The same approach is used in other approaches, including fully empirical models \cite{Mavromichalaki1984,ZhaoQin2013}.
In this respect, our formula can be regarded as a generalization of early attempts to account for the lag in \GCR modulation.

\subsection{Further developments}  
\label{Sec::FutureWork}            

The study of the relationship between \GCR modulation and solar activity
provides the basis for establishing the long-term forecast of the \GCR radiation levels in the heliosphere.
Forecasting \GCR radiation is an important concerns for crewed space travel in future missions \citep{DuranteCucinotta2011}. 
In light of the comparison with other studies, further correlative analyses may be carried out the use of other proxies
carrying different information.
Examples are indices are the open magnetic flux, the tilt angle of the heliospheric current sheet, the solar irradiance, the flare index.
Using these indices, one can in principle allows to assess the different contributions of the modulation lag.
Regarding the interpretation, we discussed our results in general terms of large-scale diffusion and drift, but 
the role of merged interaction regions in the time-dependent \GCR modulation should be investigated.
Another goal is to calculate the precise rigidity dependence of the \GCR propagation times in the difference phases of the solar cycle.
This will be done in a forthcoming work, based on stochastic simulations of \GCRs in the heliosphere and time-dependent
constraints from \AMS/PAMELA data \citep{Fiandrini2021}.

To improve the study of the rigidity dependence of NM rates, one can set up more refined calculations of the transmission function.
In this work, variations in the cutoff $R_{C}$ were neglected, although
in the last five cycles it decreased of nearly 5\,\% in the equatorial regions \citep{Cordaro2019}. 
However, NM rates remain unsuitable to study the rigidity dependence of the lag, while direct \GCR data from space would be of help.
In the present study, we considered data from ACE and IMP-8 at the GV scale.
The observational gap in the $\sim$\,1-30\,GV rigidity region may be fulfilled with
the data from new-generation experiments in low-Earth orbit such as \AMS, PAMELA and CALET.
The \AMS experiment has now covered one solar cycle of exposure. In particular, the \AMS experiment in the International Space Station
can provide measurements of \GCR antiprotons  over about one solar cycle of exposure. With time series of antimatter/matter data,
the charge-sign dependence of the modulation lag can be determined, and the physical interpretation provided here can be resolutely tested.

\section{Conclusions} 
%
In this paper, we have investigated the delayed association between the variability
of solar activity and the temporal variation of \GCR fluxes.
Using a large collection of \GCR flux measurements from spacecraft and counting rates from NMs,
we have reported the observation of important features in the solar modulation effect of \GCRs in the heliosphere.
First, we have shown that the modulation lag between SSN and \GCR flux is subjected to a quasiperiodical behavior which
follows the 22-year cycle of Sun's magnetic polarity.
Moreover, we have found that the mean value of the modulation lag decreases with the rigidity or energy of the cosmic particles.
These features reveal important aspects of the physics of \GCR modulation phenomenon.
We interpreted our findings in terms of the combination of basic processes of  charged particle transport:
drift of \GCRs over the large-scale interplanetary magnetic field, convection over the latitudinal-dependent solar wind,
and rigidity dependent diffusion in the magnetic irregularities of heliospheric turbulence.
Based on this interpretation, we have proposed an effective formula for
describing the temporal evolution and the rigidity dependence of the \GCR modulation lag.
Our formula could be used as an effective input in solar modulation models
for making predictions of the \GCR radiation driven by solar activity proxies.
Within the physical picture presented here,
the observed connection between modulation lag and solar cycle can be considered as a
remarkable signature of charge-sign dependent drift in \GCR transport. The interpretation could be resolutely
tested using data on \GCR antimatter collected from \AMS over a full decade of observation.

\section{Acknowledgements} 
%
We acknowledge the support of Italian Space Agency (ASI) under Agreement No. ASI-UniPG 2019-2-HH.0.
We also acknowledge the project \emph{DRIFT - ``Cosmic antimatter drifting through the solar wind''}
in the program Fondo Ricerca di Base 2019 of the University of Perugia.
In this work, use was made of the \emph{\href{https://www.nmdb.eu}{NMDB.eu}} real-time database (for NM rates),
the \href{https://tools.ssdc.asi.it/CosmicRays/}{Cosmic-Ray Database} of the Space Science Data Center at ASI (for \GCR data from \AMS and PAMELA),
the \emph{\href{https://omniweb.gsfc.nasa.gov}{OMNIWeb}} service of the NASA Space Physics Data Facility (for spacecraft data),
and the \emph{\href{https://www.sidc.be/silso/}{SILSO}} SSN database of the \emph{Solar Influences Data Analysis Center} at the Royal Observatory of Belgium.
We also thank Nikolay Kuznetsov and Elena Popova for support with the digitized IMP-8 data,
Ilya Usoskin for sharing the time series of the $\phi$ parameter. 


\end{document}